\documentclass[a4paper,11pt]{article}
\usepackage{jinstpub} 
\usepackage{lineno}

\usepackage[acronym]{glossaries}
\glsdisablehyper

\newacronym{tot}{ToT}{Time-over-Threshold}
\newacronym{toa}{ToA}{Time-of-Arrival}
\newacronym{tdc}{TDC}{Time-to-Digital Converter}
\newacronym{idaq}{IDAQ}{INFN Data Acquisition}
\newacronym{tlu}{TLU}{Trigger Logic Unit}
\newacronym{sc}{SC}{Slow-Control}
\newacronym{ci}{CI}{Continuous Integration}
\newacronym{pga}{PGA}{Pin Grid Array}
\newacronym{pcb}{PCB}{Printed Circuit Board}
\newacronym[plural=DAQs,firstplural=Data Acquisition systems (DAQs)]{daq}{DAQ}{Data Acquisition system}
\newacronym[plural=ASICs,firstplural=Application-Specific Integrated Circuits (ASICs)]{asic}{ASIC}{Application-Specific Integrated Circuit}
\newacronym[plural=FPGAs,firstplural=Field-Programmable Gate Arrays (FPGAs)]{fpga}{FPGA}{Field Programmable Gate Array}

\newcommand{\FourDPhoton}{4DPHOTON\ }
\newcommand{\TimepixFour}{Timepix4\ }
\newcommand{\MedipixFour}{Medipix4\ }
\newcommand{\ipbus}{IPbus\ }

\title{A flexible DAQ system for the Timepix4 ASIC in the 4DPHOTON project}

\author[a,1]{N.~V.~Biesuz\note{Corresponding author.}}
\author[c]{J.~Alozy}
\author[c]{R.~Bolzonella}
\author[a,b]{V.~Cavallini}
\author[a]{A.~Cotta~Ramusino}
\author[a,b]{M.~Fiorini}
\author[a,b]{E.~Franzoso}
\author[c]{X.~Llopart Cudie}
\author[a,b]{A.~Saputi}

 \affiliation[a]{INFN sezione di Ferrara,\\via Saragat 1 - 44122 Ferrara, Italy}
 \affiliation[b]{Universi\'a di Ferrara,\\via Ludovico Ariosto, 35 - 44121 Ferrara, ITALY}
 \affiliation[c]{CERN,\\Esplanade des Particules 1 - P.O. Box 1211 Geneva 23 Switzerland, SWITZERLAND}

\emailAdd{biesuz@fe.infn.it}

\arxivnumber{1234.56789} 

\abstract{
This paper presents the design and implementation of a flexible data acquisition system developed for the Timepix4 ASIC. 
The system is based on a modular FPGA-centric architecture combining high-speed serial readout, Ethernet-based data transport, and deterministic multi-board synchronization. 
The hardware platform includes a scalable stack composed of commercial FPGA carrier boards, FMC-based interface electronics, detector-specific chipboards, and a dedicated Trigger Logic Unit for synchronized operation. 

The firmware architecture separates control and data paths, enabling independent configuration and high-throughput acquisition through UDP over 10~GbE links. 
The design supports zero-back-pressure operation toward the ASIC and allows adaptation of the readout bandwidth to different experimental conditions. 
Synchronization between multiple DAQ systems is achieved through a common clock and trigger distribution network, experimentally demonstrated with sub-100~ps precision.

The system has been developed to support detector characterization, laboratory measurements, and beam-test campaigns for the 4DPHOTON detector concept. 
Hardware organization, firmware architecture, synchronization strategy, and performance measurements are presented.
}


\begin{document}
\maketitle
\flushbottom
\section{Introduction}\label{sec:introduction}

\begin{figure}[htb]
\centering
\includegraphics[width=0.55\textwidth]{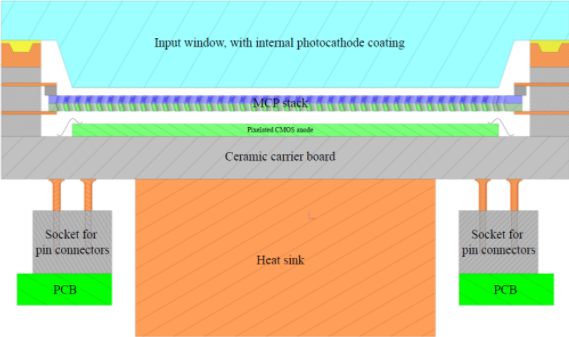}
\caption{Schematic cross section view of the \FourDPhoton detector, taken from~\cite{Alozy2022SinglePhoton}.}
\label{fig:tube_cross_section}
\end{figure}

The \FourDPhoton project aims at the development of a single-photon imaging detector combining high spatial granularity with picosecond-level timing capability.
The detector concept follows a hybrid architecture in which a photocathode converts incident photons into electrons, a microchannel plate (MCP) stack provides gain through secondary emission, and a pixelated anode \gls{asic} performs charge collection and time stamping~\cite{Fiorini2018,Alozy2022SinglePhoton}.
A schematic cross-section is shown in Figure~\ref{fig:tube_cross_section}.

The readout anode is based on the \TimepixFour ASIC, a 55\ $\mu$m pitch hybrid pixel detector featuring a four-side buttable architecture and sub-200~ps time binning capability~\cite{Llopart2022Timepix4}.
In data-driven mode, Timepix4 provides simultaneous \gls{toa} and \gls{tot} measurements for each pixel hit with an intrinsic time resolution below 100~ps~rms for sufficiently large input signals, while sustaining hit rates up to $\sim$3.6~Mhits/mm$^2$/s.
Its analog front-end, optimized for low noise and low jitter operation in 65~nm CMOS technology, achieves an equivalent noise charge of $\sim$65--80~e$^-$~rms depending on gain configuration~\cite{Alozy2022SinglePhoton}.

Within this framework, a dedicated \gls{daq} has been developed to interface the \TimepixFour and to allow future extension to the \MedipixFour \gls{asic}. The system emphasizes modularity, cost-effectiveness, and ease of deployment in laboratory and beam-test environments, relying on standard high-speed serial links and Ethernet-based data transport to satisfy the timing and throughput requirements of the project.

\section{The \TimepixFour ASIC}\label{sec:tpx4_mpx4_asics}

The Medipix and Timepix family\cite{Ballabriga2018} consists of hybrid pixel readout \glspl{asic} developed for radiation imaging and particle detection. By combining low input capacitance ($10-100\ \mathrm{fF}$) with on-pixel signal processing, these devices achieve low noise performance and precise charge measurement in hybrid pixel detectors.

Two complementary architectural approaches define the family. The Medipix line targets photon-counting and spectroscopic imaging using a frame-based readout in which counts are accumulated locally in each pixel. The Timepix line focuses on transmitting detailed information about individual interactions, enabling timing and energy-related measurements at the pixel level.

The most recent device of the Timepix family, Timepix4, extends these concepts. \TimepixFour provides data-driven readout with simultaneous \gls{toa} and \gls{tot} information, optimized for high-rate, time-resolved applications. 


\subsection{\TimepixFour}\label{sec:tpx4}

The \TimepixFour\ \gls{asic} is the latest-generation device in the Timepix family, developed by the Medipix collaboration in a 65~nm CMOS technology. 
It features a four-side buttable matrix of $448 \times 512$ pixels with a $55~\mathrm{\mu m}$ pitch, enabling seamless large-area tiling~\cite{Llopart2022Timepix4,Ballabriga2023Timepix4AFE}. 
Each pixel integrates a charge-sensitive front-end, local discriminator, and a \gls{tdc}, allowing simultaneous measurement of \gls{toa} and \gls{tot} for signals above a programmable threshold~\cite{Llopart2022Timepix4}.

In data-driven mode, the chip outputs 64-bit event words only for hit pixels, with $195~\mathrm{ps}$ timestamp binning and support for hit rates of several $\mathrm{MHz/mm^2}$~\cite{Llopart2022Timepix4,Ballabriga2023Timepix4AFE}. 
The analog front-end is configurable to optimize gain and noise performance, achieving equivalent noise charges of a few tens of electrons rms depending on configuration~\cite{Ballabriga2023Timepix4AFE}. 
Data are serialized over up to 16 high-speed links, each programmable between $40~\mathrm{Mb/s}$ and $10.24~\mathrm{Gb/s}$, providing sufficient aggregate bandwidth for high-rate operation.

If maximum rate performance is not required, the off-chip electronics can be simplified by enabling only the number of serial links needed, as well as limiting the link data rate, required by the expected data rate.

Each hit produces a 64-bit word transmitted using 64b/66b encoding with a self-synchronizing scrambler defined by the polynomial $X^{58}+X^{39}+1$, corresponding to 66 transmitted bits per hit.

Assuming a uniform hit distribution, the maximum sustainable average hit rate per pixel is

\[
hit\_rate \,\left[\frac{\mathrm{hits}}{\mathrm{s}}\right] =
\frac{n_{\mathrm{links}} \cdot link_{\mathrm{speed}}}
{66 \cdot 256 \cdot 448}
\]

where $n_{\mathrm{links}}$ is the number of active multi-gigabit links per half-matrix, $link_{\mathrm{speed}}$ the selected line rate in $\mathrm{bit/s}$, and $256 \times 448$ the number of pixels in one half-matrix.

Data acquisition is controlled by two signals, \texttt{T0\_SYNC} and \texttt{SHUTTER}. 
\texttt{T0\_SYNC} resets the global time reference, while \texttt{SHUTTER} enables data taking, allowing signals from the bump-bonding pads to propagate to the readout only when asserted.

Configuration and monitoring are handled via a dedicated Slow Control interface (Section~\ref{sec:sc}).

These features make \TimepixFour\ well suited for hybrid pixel detectors requiring high spatial resolution, per-event timing, and large throughput, as in the \FourDPhoton\ detector concept.

\subsection{Slow Control interface}\label{sec:sc}

The \TimepixFour \gls{asic} is configured through a dedicated Slow Control interface common to \TimepixFour and Medipix4. 
The interface consists of two unidirectional LVDS buses, one for input and one for output, each comprising a serial data line and a serial clock line.

The protocol follows a Master–Worker scheme, where the off-detector electronics acts as Master and the \gls{asic} as Worker. 
Communication is initiated by the Master through a command packet containing a custom header, an operation field (read/write), the target register address, and optional data. 
The Worker decodes the packet, executes the requested operation, and returns a reply packet including a header, status bits, the register address, and optional data.

Error conditions are reported via dedicated reply packets, which must be handled by the off-chip electronics.

The exact packet length depends on the register map and the requested operation, as registers may have different data widths and payload sizes. 
To enable reuse of a common configuration middleware, the device-specific register mapping is abstracted, as described in Section~\ref{sec:fw_tpx4}.

\section{\TimepixFour DAQ system: state of the Art}

Several FPGA-based DAQ platforms have been developed for Medipix and Timepix ASICs. 
The SPIDR (Speedy PIxel Detector Readout) system was originally designed at NIKHEF for Medipix3 and Timepix3. 
Its successor, SPIDR4~\cite{Heijden2017}, targets \TimepixFour and supports up to 16 serial links operating at 10~Gb/s, enabling aggregate bandwidths of approximately 160~Gb/s. 
Reaching this maximum bandwidth requires dedicated control and DAQ server hardware in addition to the front-end board.
If no such hardware is present the total output data bandwidth is limited to $5.12\ \mathrm{GB/s}$.
While this architecture is well suited for high-rate applications, such throughput is often unnecessary in small-scale setups, such as laboratory measurements or test-beam campaigns, where the rate requirements are significantly lower, yet the lower bandwidth can generate back pressure also in these settings.

The Katherine readout represents an alternative \TimepixFour DAQ targeted primarily at laboratory and test-beam applications~\cite{KatherineTPX4}. 
It supports up to four \TimepixFour detectors and eight serial links, with data transfer through either 1~GbE or PCIe Gen3~$\times$4. 
The latter provides a maximum reported throughput of approximately $350~\mathrm{Mhits/s}$.
The system also performs low-level processing of \TimepixFour data directly in hardware and provides a modular software architecture.

Other available \TimepixFour DAQ implementations follow a similar high-throughput, data-driven approach optimized specifically for the Timepix4 protocol. 
However, these systems are tailored to the data-driven architecture of \TimepixFour and are not easily extensible to the Medipix family. 
This motivates the development of a unified and scalable DAQ framework capable of addressing both ASIC families while remaining adaptable to different rate regimes.

\section{IDAQ Hardware system description}\label{sec:idaq_arch}

\begin{figure}[htb]
\centering
\includegraphics[width=0.75\textwidth]{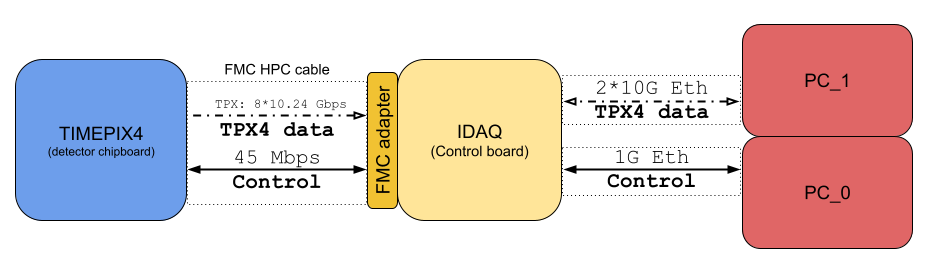}
\caption{Top-level architecture of the \gls{daq} system.}
\label{fig:system}
\end{figure}

The \gls{idaq} architecture is based on a modular hardware stack, shown in Figure~\ref{fig:system}, composed of:
\begin{itemize}
  \item an \gls{idaq} control board;
  \item an \gls{idaq} FMC adapter card;
  \item an \gls{asic}-specific chipboard hosting the detector;
  \item a custom \gls{tlu}.
\end{itemize}

The system interfaces the front-end electronics to remote servers. 
The control board (Section~\ref{sec:idaq}) connects to the servers through one 1~GbE link for configuration and monitoring and two 10~GbE links dedicated to detector data transmission. 
Separating control and data paths enables the use of dedicated servers for configuration/monitoring and high-rate data processing, allowing computationally intensive tasks to be offloaded. 
When required, both services can also run on a single server. 
The two 10~GbE links provide an aggregate output bandwidth of more than $ 16~\mathrm{Gb/s}$, including UDP/IP overhead, which is sufficient for the targeted applications.

The detector is connected via the FMC high-pin-count connector on the control board. 
Although direct interfacing to the \gls{asic}-specific chipboard is possible, an intermediate FMC adapter (Section~\ref{sec:fmc_adapter}) is provided. 
This board ensures compatibility with existing chipboards that do not fully comply with the FMC standard, using dedicated single-ended and LVDS buffers. 
It also integrates a clock management stage to synchronize multiple \gls{idaq} systems operating in parallel.

The \gls{asic}-specific chipboard hosts the detector assembly, distributes power to the front-end electronics and sensor, and exposes the required control signals to the \gls{daq}. 
While compatible with chipboards developed within the collaboration, a dedicated board (Section~\ref{sec:tube_card}) was designed in the \FourDPhoton project to support tube production tests, characterization, and test-beam campaigns.

\subsection{FPGA carrier board}\label{sec:idaq}

The FPGA carrier board, referred to as the \gls{idaq} control board, is based on a commercial development kit. 
The selected platform is the \textit{AMD Kintex UltraScale FPGA KCU105 Evaluation Kit} (EK-U1-KCU105-G), chosen because it provides all required resources for the \gls{idaq} implementation.
Migration toward the lower-cost \textit{Alinx AXKU040} platform is ongoing and will retain all the system features while lowering the system cost.

As described in Section~\ref{sec:idaq_arch}, a single 1~GbE link is used for configuration and monitoring. 
On the KCU105 board this connection is implemented through an RJ45 connector interfaced to the FPGA via a Marvell M88E1111 external PHY. 
The PHY communicates with the FPGA using an SGMII data interface and an MDIO management bus.

High-speed data transmission is handled by two 10~GbE links implemented through SFP+ cages directly connected to the FPGA transceivers. 
A single FMC high-pin-count (HPC) connector is available for detector interfacing. 
This connector supports up to eight multi-gigabit serial lines, which constrains the number of \TimepixFour\ output links that can be received simultaneously.

The board hosts an \textit{AMD Kintex UltraScale XCKU040-2FFVA1156E} \gls{fpga}. 
A summary of its main resources is reported in Table~\ref{tab:KCU040_summary}. 
The firmware implementation for this device is described in Section~\ref{sec:fw_tpx4}.

\begin{table}
    \centering
    \begin{tabular}{cc}
         Resource Type & Resource count \\
         \hline
         System Logic Cells & $530 000$ \\
         DSP Slices & $1 920$ \\
         Block RAM & $21.1\ Mb$\\
         16.3Gb/s Transceivers & $20$ \\
         I/0 Pins & $520$\\
    \end{tabular}
    \caption{AMD Kintex UltraScale XCKU040 resource summary table}
    \label{tab:KCU040_summary}
\end{table}

\subsection{FMC adapter}\label{sec:fmc_adapter}

\begin{figure}[htb]
    \centering
    \begin{minipage}{0.49\textwidth}
        \centering
        \includegraphics[width=0.75\linewidth]{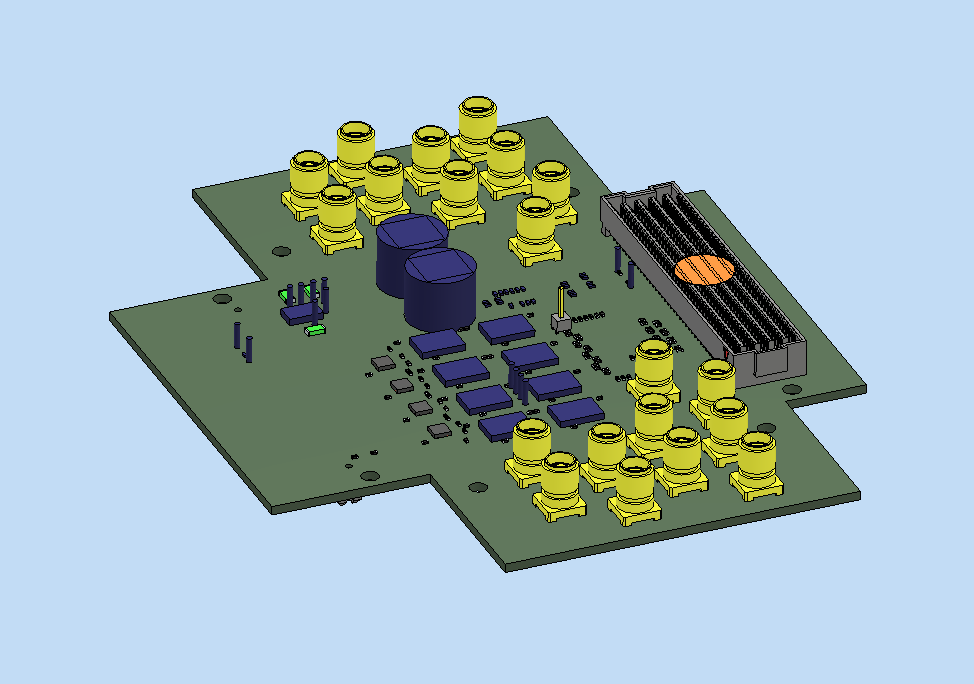}
    \end{minipage}
    \hfill
    \begin{minipage}{0.49\textwidth}
        \centering
        \includegraphics[width=0.75\linewidth]{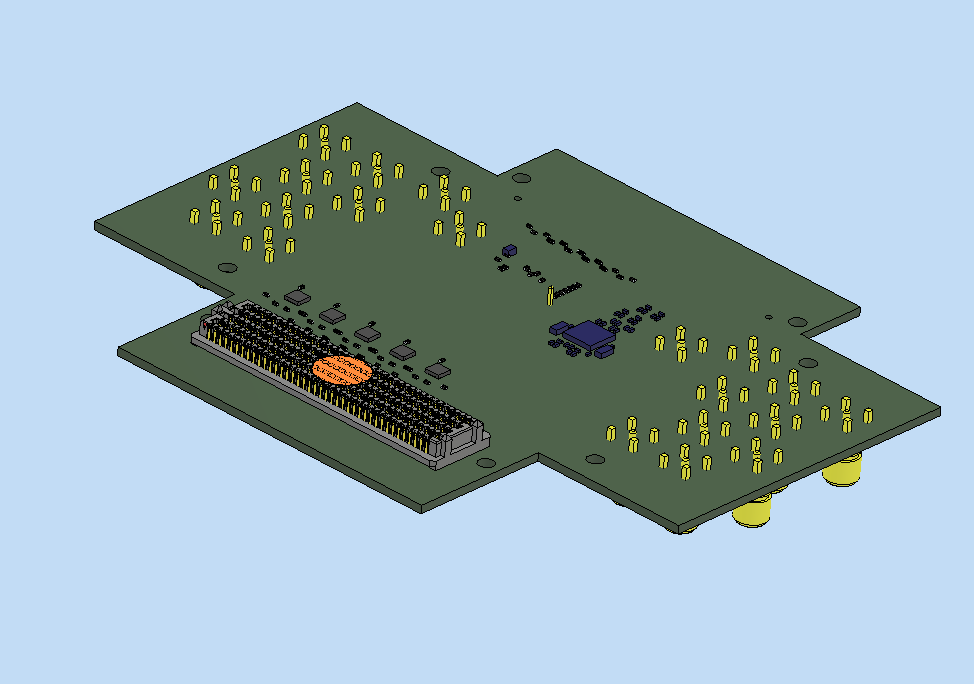}
    \end{minipage}
    \caption{3D rendering of the FMC adapter card assembly viewed from the top side (left) and bottom side (right).}
    \label{fig:fmc_adapter}
\end{figure}

The FMC adapter, shown in Figure~\ref{fig:fmc_adapter}, provides electrical and protocol adaptation between the FPGA carrier and the \gls{asic} chipboard. 
Its functions include level translation, buffering of high-speed links, clock distribution, and routing of control signals. 

Single-ended signals are translated from the $1.8~\mathrm{V}$ LVCMOS standard used by the FPGA to the $3.3~\mathrm{V}$ logic level required by the chipboards using bidirectional level translators. 
These devices support both unidirectional TTL control signals (e.g. power-enable and power-good) and bidirectional open-drain protocols such as I\textsuperscript{2}C.

Differential control signals are buffered using re-drivers, compatible with the SLVS transmitter standard employed by the \TimepixFour\ \gls{asic}. 

Additional buffering is provided for the \TimepixFour data outputs, which use SSTL transmitters operating between $40~\mathrm{Mb/s}$ and $10.24~\mathrm{Gb/s}$. 
In this project, data rates of $1.28~\mathrm{Gb/s}$, $2.56~\mathrm{Gb/s}$, and $5.12~\mathrm{Gb/s}$ are targeted. 
A Diodes Incorporated PI3EQX12908A eight-channel differential re-driver is used to compensate transmission losses when the detector is located remotely from the control board. 
This equalization is necessary because the \TimepixFour\ transmitters do not provide programmable output equalization. 
The re-driver is configured through a dedicated I\textsuperscript{2}C interface.

\begin{figure}[htb]
\centering
\includegraphics[width=0.75\textwidth]{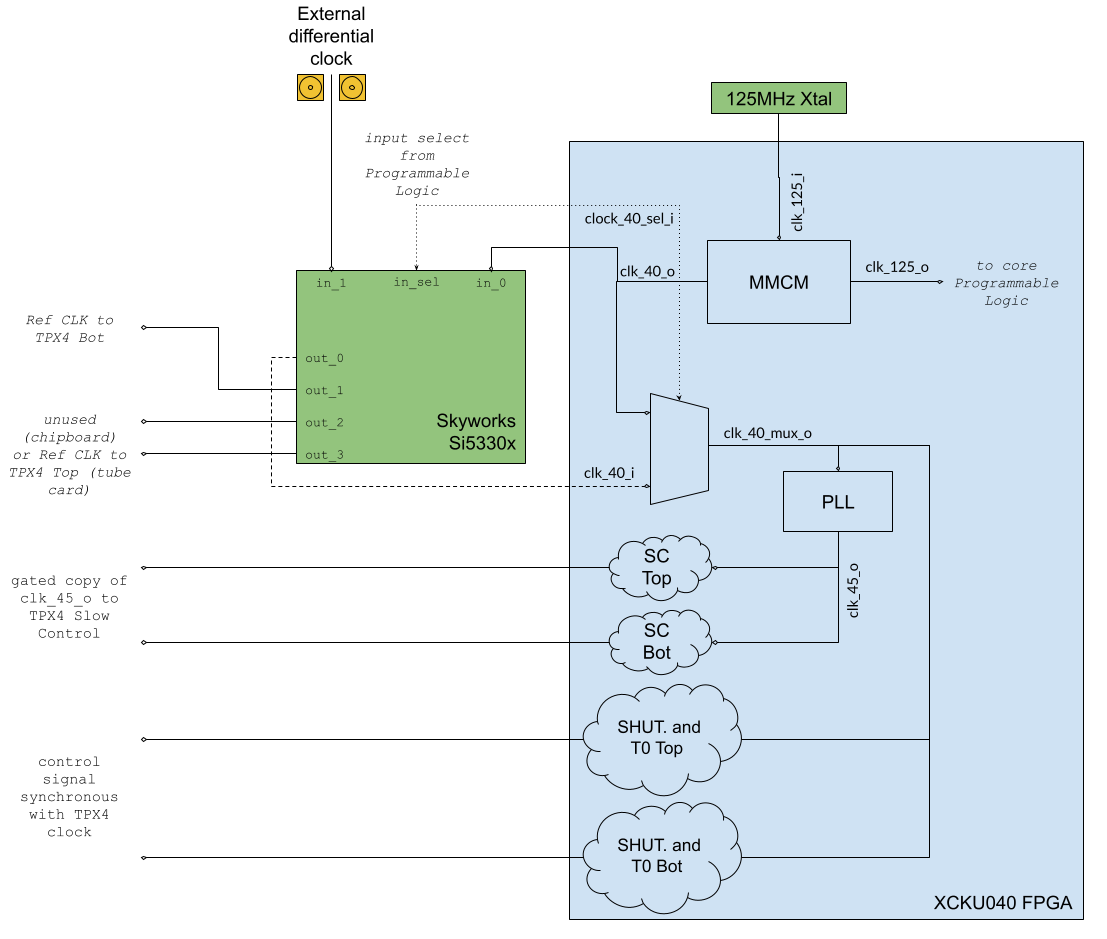}
\caption{Schematic view of the clock synchronization mechanism}
\label{fig:clk_sync}
\end{figure}

The FMC adapter also implements clock synchronization. 
A simplified schematic is shown in Figure~\ref{fig:clk_sync}. 
A 1 to 4 clock buffer receives two $40~\mathrm{MHz}$ clock inputs: a default clock generated by the FPGA and an external clock provided via SMA connectors for multi-system synchronization. 
By default, the FPGA clock is forwarded to the chipboard. 
The FPGA monitors the Loss-of-Signal outputs of the clock buffer inputs and, when multi-system synchronization is enabled and a valid external clock is detected, switches the forwarded clock to the SMA source. 
The selected clock is also routed back to the FPGA, ensuring synchronous communication between the FPGA logic and the \TimepixFour\ \gls{asic}.

\subsection{Chipboard design}

The chipboard hosts the \TimepixFour\ \gls{asic} and provides power distribution, configuration interfaces, and high-speed data connectivity. 
Chipboards for the Timepix4, bump-bonded to solid-state pixelated sensors, were developed by collaborating institutions and used for system validation.

The \TimepixFour chipboard produced by NIKHEF is supported through the optional FMC adapter card. 
The \FourDPhoton tube chipboard was designed to be pin-compatible with this NIKHEF board, simplifying development while ensuring compatibility with the existing SPIDR4 \gls{daq} system. 
This approach allowed reuse of proven electrical interfaces and reduced integration risks.


\section{Chipboard for the \FourDPhoton tube}\label{sec:tube_card}

The chipboard design was adapted for the \TimepixFour integration inside the \FourDPhoton detector tube, which introduces additional mechanical, thermal, and electrical constraints.

\subsection{\FourDPhoton ceramic carrier and tube socket}

\begin{figure}[htb]
    \centering
    \begin{minipage}{0.49\textwidth}
        \centering
        \includegraphics[width=0.75\linewidth]{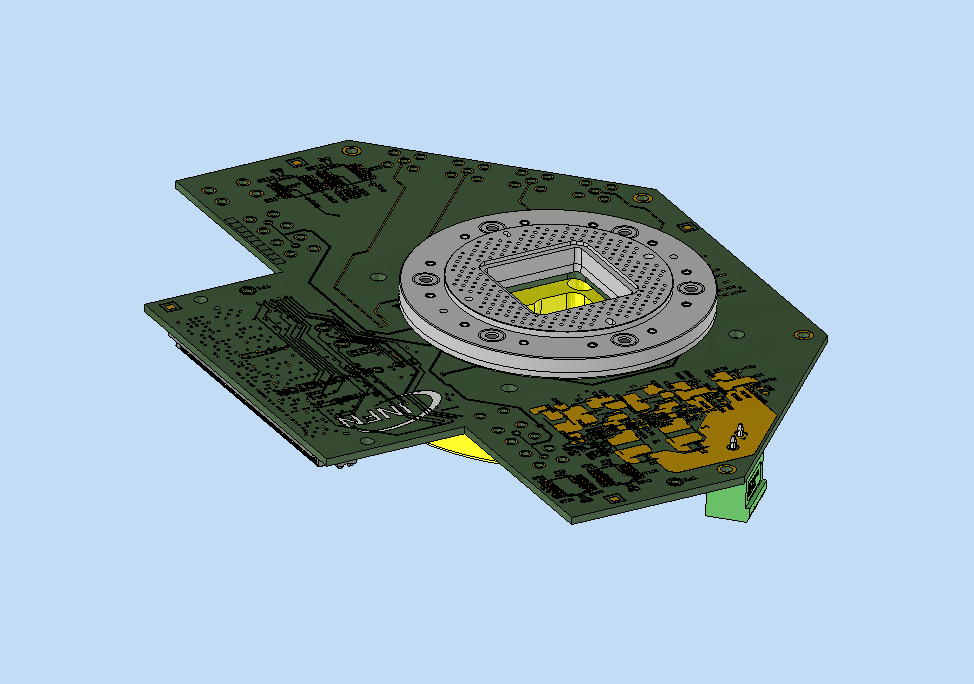}
    \end{minipage}
    \hfill
    \begin{minipage}{0.49\textwidth}
        \centering
        \includegraphics[width=0.75\linewidth]{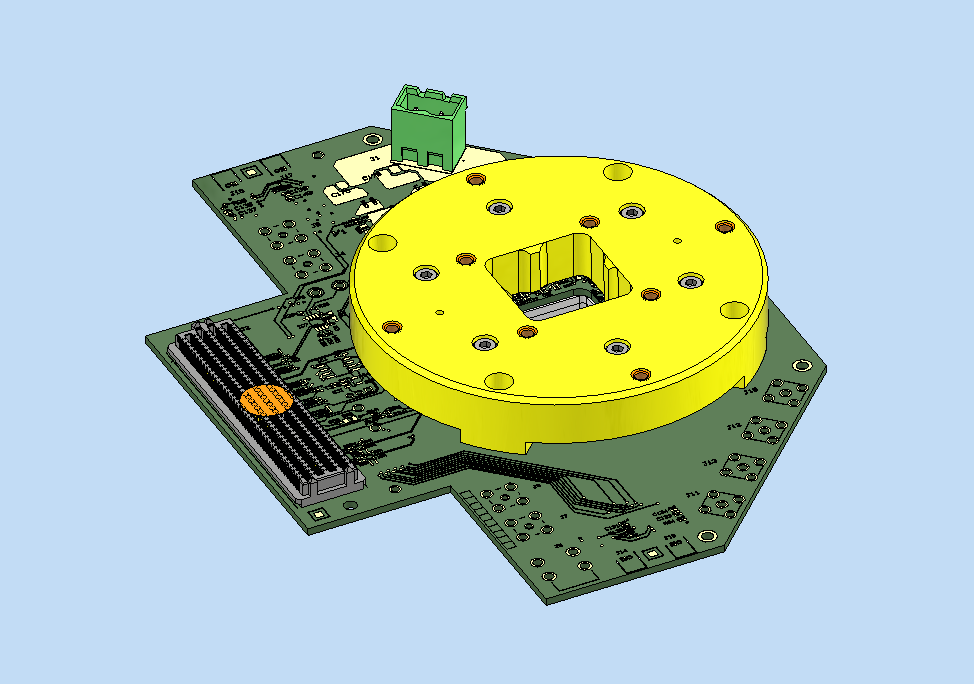}
    \end{minipage}
    \caption{3D rendering of the tube card assembly viewed from the top side (left) and bottom side (right).}
    \label{fig:tube_card}
\end{figure}

The \FourDPhoton tube chipboard (Fig.~\ref{fig:tube_card}), namely the tube card, interfaces the front-end ASIC through a custom ceramic carrier providing electrical routing and thermal coupling. 
The carrier is a cylindrical 35-layer ceramic \gls{pcb} with a thickness of approximately $8.2~\mathrm{mm}$ and a diameter of about $43~\mathrm{mm}$. 
Its geometry is defined by vacuum compatibility and mechanical constraints imposed by the tube manufacturer.

The bare \TimepixFour ASIC is mounted on the top layer and attached to a grounded pad to ensure efficient heat transfer. 
Custom wire-bond pads provide electrical connectivity while maintaining a low bonding profile, enabling proximity focusing of the electron cloud from the micro-channel plate.

Power and data signals are routed to the bottom layer, where a 204-pin \gls{pga} with $9~\mathrm{mm}$ pins interfaces to the downstream electronics. 
A central $20 \times 20~\mathrm{mm^2}$ opening accommodates a custom heatsink.

Electrical contact between the PGA and the chipboard is realized through a custom socket based on spring-loaded pogo pins embedded in a plastic matrix. 
A central aperture in the socket allows insertion of the heatsink to provide cooling for the \TimepixFour.

\subsection{Tube card design}

The tube card hosts the socket and mechanical clamps securing the \FourDPhoton tube. 
A central opening in the \gls{pcb} aligns with the tube socket aperture to allow heatsink insertion. 
The board outline enables up to four tube cards to be arranged in a circular configuration, matching the foreseen use of \FourDPhoton tubes in Ring Imaging Cherenkov detectors, where the Cherenkov ring spans four sensors.

Particular attention was devoted to low-noise power delivery for the \TimepixFour. 
An external linear supply provides $2~\mathrm{V}$ to two pairs of on-board linear regulator stages generating the required $1.2~\mathrm{V}$ rails. 
The first pair supplies the high-current analog and digital core domains using Texas Instruments TPS7A5701RTER LDO regulators rated up to $5~\mathrm{A}$ each, exceeding the maximum specified consumption of $<3.5~\mathrm{A}$ (analog) and $<2.5~\mathrm{A}$ (digital). 
A Schottky diode at the regulator input reduces the effective input voltage and protects against reverse bias. 
LC filters at both input and output suppress regulator-induced noise.

The second regulator pair powers the multi-gigabit transceivers and PLLs using Analog Devices LT3045 LDOs rated at $500~\mathrm{mA}$, again with input and output LC filtering.

Decoupling capacitors are distributed on all power rails and placed on the bottom layer beneath the tube socket region. 
Due to mechanical constraints, these components are located approximately $20~\mathrm{mm}$ from the ASIC power entry point; therefore, extensive local decoupling and dedicated power split-planes are employed to minimize impedance and maintain supply stability.

\section{Firmware for \TimepixFour}\label{sec:fw_tpx4}

\subsection{Firmware distribution and versioning}\label{sec:hog}

The firmware distribution and versioning schema will be discussed first since this will introduce concepts used in the firmware architecture description.

The firmware is hosted in a Gitlab repository containing the source files and recipes for generating the compilable AMD Vivado projects.
Strict versioning ensures reproducibility and traceability of firmware source files.

Generation of the AMD Vivado project is performed using Hog\cite{Hog2023} tool.
This tool provided scripts for compilation of the firmware in gitlab \gls{ci}\cite{GitLabCI} and locally, ensuring each release can be traced back to the exact source files used for production.
The firmware can be distributed in the form of gitlab releases containing all relevant information.

Development follows a standard git workflow with a stable main branch and feature-specific development branches. 
When a feature is completed, a merge request triggers a GitLab \gls{ci} pipeline executing Hog routines that generate a clean project copy, perform synthesis and implementation, execute self-checking tests, and build the firmware documentation. 
If all steps complete successfully, the merge request can be accepted, merging the feature into the main branch and automatically generating a tagged firmware release.
The release contains the compiled firmware, a version table for the firmware, all compilation reports, any helper scripts provided by the developer (e.g. scripts to program the device, check the firmware versions), a ChangeLog, the generated documentation and a memory map of the \gls{fpga} address space in the form of xml files.

Firmware releases are distributed as Open Hardware under the CERN Open Hardware Licence to the collaboration; an additional automated validation stage based on the REUSE software~\cite{REUSEsoftware} verifies the presence of the correct copyright and license information in all source files.

\subsection{Firmware architecture}\label{sec:fw_arch}

The firmware for the XCKU040 \gls{fpga} has been developed around the idea of maintaining the control and data paths independent.

\begin{figure}
    \centering
    \includegraphics[width=0.75\linewidth]{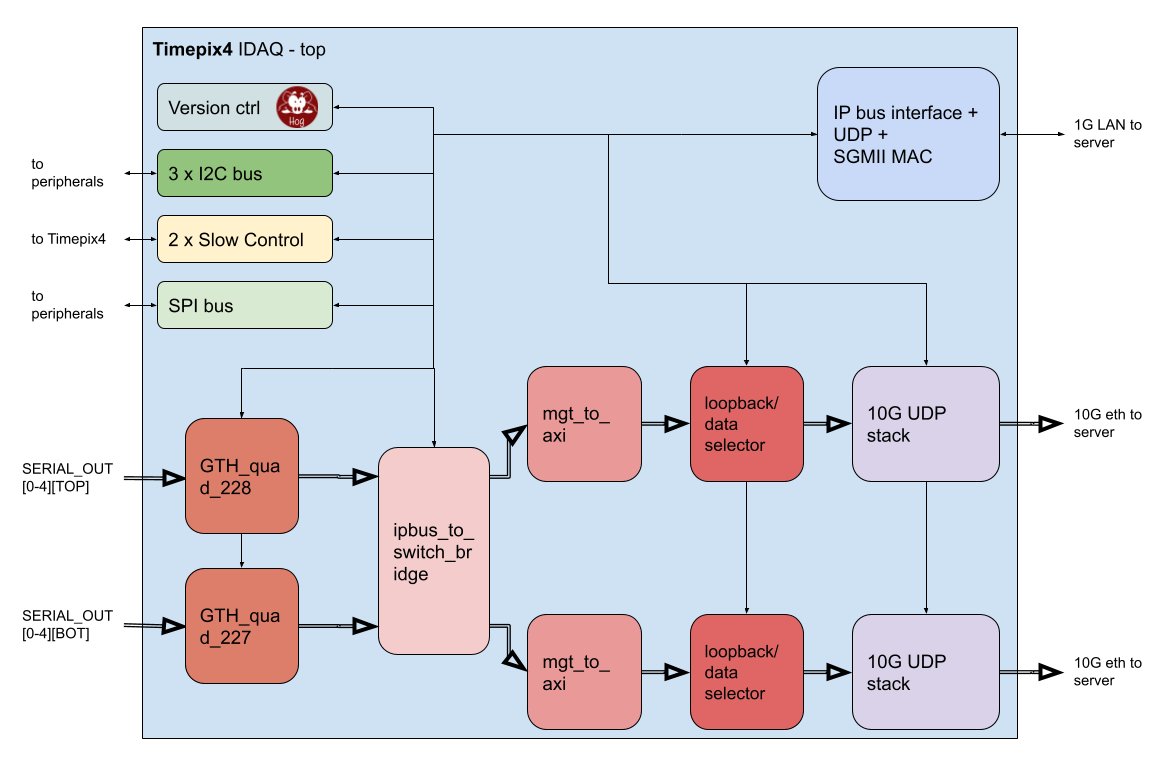}
    \caption{IDAQ firmware schema. The schema reports the top level architecture for the project. Instances of the same entity are highlighted using the same colour. solid single arrows indicate the control path, solid double arrows indicate the high speed data path.}
    \label{fig:cbf_top}
\end{figure}

The control path, indicated by single arrows in Figure~\ref{fig:cbf_top}, is built around the \ipbus protocol~\cite{IPbus2015}.
This protocol allows to expose the \gls{fpga} configuration space over the 1G ethernet connection.

The data path, indicated by double arrows in Figure~\ref{fig:cbf_top}, is instead thought to be a zero-back-pressure path connecting the \TimepixFour data output for a single half-matrix to one 10G ethernet connection.
Two exact copies of this path are present, one per half-matrix, thus exploiting the two 10G ethernet connection on the board.
Thus the control path interacts with this path during configuration but any interaction is frozen during data-taking. 

\subsection{Control path}\label{sec:control_path}

The \ipbus protocol exposes the \gls{fpga} configuration space over the 1~GbE link. 
The AMD Tri-Mode Ethernet MAC (TEMAC) IP, interfaced to the external PHY through a GMII–SGMII bridge, forwards UDP/IP frames to the \texttt{ipbus\_ctrl} module via AXI. 
These frames are decoded into \ipbus transactions and transmitted on a dedicated control bus.

The \gls{fpga} address space follows a hierarchical structure grouping registers by module. 
Address-to-function mappings are defined in XML files distributed with each firmware release and required by the control software. 
During GitLab \gls{ci} project generation, expected version register values are embedded in the XML to guarantee firmware–software consistency. 
The same XML description is used to automatically generate VHDL packages implementing the address mapping; these generated files are excluded from version control to avoid conflicts. 
The hardware mirrors this hierarchy by splitting the \ipbus control bus into sub-buses corresponding to XML levels, using helper functions from the generated packages. 

Firmware modules (VHDL-2008) consist of a functional core and an \ipbus bridge exposing registers and memories. 
The implemented modules include:
\begin{itemize}
    \item version and status register bank (firmware versions and board monitoring);
    \item board configuration register banks;
    \item I\textsuperscript{2}C and SPI bridges (system bus, FMC re-driver, secondary \TimepixFour configuration, temperature monitoring);
    \item slow control bridge (\gls{asic} configuration);
    \item data-path modules:
    \begin{itemize}
        \item 10G Ethernet configuration registers;
        \item transceiver configuration registers for \TimepixFour inputs;
        \item data-switch bridge routing transceivers to 10G outputs.
    \end{itemize}
\end{itemize}

\subsection{Data path}\label{sec:data_path}

The \TimepixFour data path is designed to operate without propagating back-pressure to the \gls{asic}; consequently, the average input data rate must be compatible with the available output bandwidth. 
Back-pressure between internal data-path modules is handled within the FPGA through appropriate buffering, but it is never reflected upstream toward the ASIC. 
As a result, instantaneous input rates exceeding the sustained output bandwidth can be tolerated for short intervals, on the order of a few microseconds, until internal buffers are saturated.

Although the data path is configured through the control path, all interactions between the two are disabled once the data-path modules are enabled. 
This strict separation prevents unintended configuration transactions from interfering with the data stream and ensures data integrity during operation.

The maximum output bandwidth is determined by the two 10G Ethernet interfaces and the associated 10G UDP stack\footnote{In the following, the complete 10G Ethernet UDP–IP–MAC–PHY stack is referred to as the 10G UDP stack for brevity.}. 
For this project, an open-source \textit{verilog-ethernet} library~\cite{VerilogEthernet} was adopted. 
The library implements UDP/IP transmission compliant with IEEE~802.3~\cite{IEEE8023}, which limits the Ethernet frame payload (MTU) to 1500 bytes, corresponding to a maximum UDP payload of 1472 bytes.

A dedicated module, \texttt{mgt\_to\_axi}, subdivides the continuous data stream from the \TimepixFour\ \gls{asic} into 1472-byte blocks. 
This module also converts the native transceiver interface (data + \texttt{data\_valid}) into the AXI-Stream protocol required by the 10G UDP stack. 
The state machine operates synchronously with the $156.25~\mathrm{MHz}$ UDP clock and introduces a latency of three additional clock cycles between consecutive 1472-byte blocks. 
It is designed to forward data to the UDP stack even when a complete 1472-byte block is not available avoiding data retention in the FPGA buffers for long time periods.

Multiple 512-word deep, 64-bit wide FIFOs are placed at the input and output of data-path modules, providing buffering equivalent to approximately 10 full 1472-byte blocks. 
Given the complexity of the UDP interaction, a dedicated hardware validation test was implemented; the results are presented in Section~\ref{sec:performance}.

An \texttt{ipbus\_to\_switch\_bridge} module is connected to the input of the \texttt{mgt\_to\_axi} block. 
In its default configuration, this module receives the data from up to eight \TimepixFour output links routed to the \gls{fpga} and forwards a single selected link to the \texttt{mgt\_to\_axi} input.

Two implementations are provided. 
The first connects a single \TimepixFour link operating at $2.56~\mathrm{Gb/s}$ to the output. 
The selected link is configured via the control path and remains fixed during data taking. 
In this configuration, the data path is purely combinatorial and cannot introduce back-pressure or data loss.

The second implementation merges data from four \TimepixFour links operating at $2.56~\mathrm{Gb/s}$ into a single output stream. 
Zero back-pressure is ensured by introducing buffering within the \texttt{ipbus\_to\_switch\_bridge} and by discarding \TimepixFour control words replicated across all active links. 
If congestion occurs, data words are dropped; this mechanism is monitored through dedicated status registers.
A true zero-back-pressure data-path can be obtained also in this configuration by masking one or more \TimepixFour output links.
This mechanism also allows to tailor the output bandwidth to the application.
If on the one hand short burst of high density hits are foreseen then all the available \TimepixFour links can be enabled.
In this case the peak data rate is adsorbed by buffers in the FPGA.
On the other hand if a constant data rate is foreseen, one link per periphery can be disabled reducing the peak data rate below the maximum sustained output data rate.

This module is the only block in which multiple data streams are merged into a single path. 
Its architecture enables application-specific customization of the \gls{idaq} system, allowing the introduction of tailored data-reduction algorithms within the merging stage. 

The \texttt{ipbus\_to\_switch\_bridge} module receives data from two quads of GTH transceivers. 
The GTH blocks are encapsulated within a dedicated wrapper entity that implements the complete transceiver initialization sequence, block-lock detection, 64b/66b de-scrambling, and clock-domain crossing from the GTH parallel clock domain to the $156.25~\mathrm{MHz}$ UDP processing clock domain.
At this stage, additional buffering is provided by using two 512-word deep and 64-bit wide FIFOs for each \TimepixFour link. This buffering mechanism mitigates rate fluctuations introduced by packetization and flow-control effects within the UDP/IP stack. 
Predefined GTH configurations are provided for line rates of $1.28~\mathrm{Gb/s}$, $2.56~\mathrm{Gb/s}$, and $5.12~\mathrm{Gb/s}$.

\section{Trigger Logic Unit}\label{sec:tlu}

The \gls{tlu} generates the global synchronization signals distributed to the \gls{idaq} boards and associated front-end electronics. 
It is implemented using an \textit{AMD Kintex UltraScale FPGA KCU105 Evaluation Kit} coupled to a custom PCB connected to the FMC-HPC interface.

The synchronization scheme is based on distributing identical copies of the \texttt{T0\_SYNC} and \texttt{SHUTTER} signals, synchronous to an external $40~\mathrm{MHz}$ reference clock. 
This clock is shared between the \gls{tlu} and all connected \gls{idaq} boards, ensuring deterministic timing alignment throughout the acquisition chain.

\begin{figure}[htb]
    \centering
    \includegraphics[width=\linewidth]{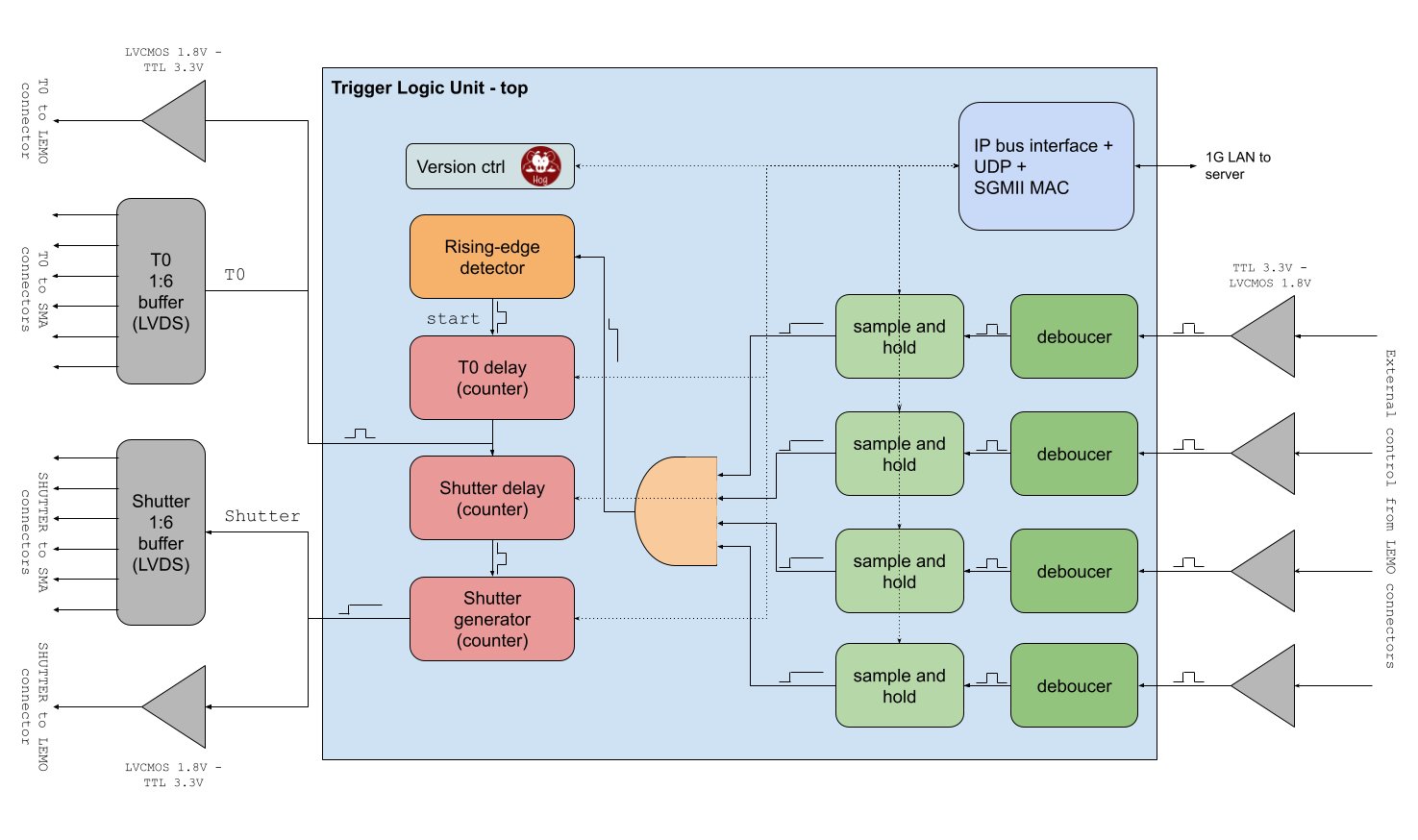}
    \caption{Schematic view of the TLU logic and hardware.\\
    Hardware components are indicated in gray. The blue square indicates the firmware logic. Rounded blue squares indicate the control and versioning logic, green modules indicate the input logic, orange components indicate the core logic and red squares indicate the output logic. Different shades indicate instances of the same entity.}
    \label{fig:TLU_schema}
\end{figure}

Figure~\ref{fig:TLU_schema} shows a schematic view of the TLU logic and hardware.
Signal generation can be triggered by up to four external inputs routed to the FPGA through dedicated buffers on the custom PCB. 
These inputs may originate from accelerator timing signals (e.g., warning-of-extraction) or from external trigger detectors. 
Unused inputs are masked in firmware by forcing the sample and hold logic to output '1'. 
Active inputs are de-bounced and stretched according to set timing parameters in the sample and hold logic and combined through a logical \texttt{AND}. 
The rising edge of this combined signal initiates a new acquisition sequence asserting the \texttt{T0\_SYNC} and  \texttt{SHUTTER} signals.
Both signals feature configurable delay and SHUTTER-pulse width parameters to adapt to different experimental conditions.

For each control signal, two electrically independent outputs are generated: a differential LVDS version and a single-ended CMOS version. 
The LVDS copy is fanned out using 1:6 differential buffers and is used for detector synchronization, providing low skew and high signal integrity. 
The single-ended copy, converted to $3.3~\mathrm{V}$ TTL levels outside the FPGA, is intended for synchronization of auxiliary or external devices.

This architecture ensures coherent multi-board synchronization while maintaining flexibility for integration with external timing and trigger systems.

\section{Performance evaluation}\label{sec:performance}

Performance characterization includes measurements of sustained throughput and synchronization accuracy.
These metrics validate the suitability of the \gls{idaq} system for high-rate timing applications.




\subsection{Firmware performance}

\textbf{Datapath sustainable data rate} - As discussed in Section~\ref{sec:data_path}, a custom state machine is inserted upstream of the UDP stack to enforce the 1500-byte MTU constraint of the IEEE~802.3 standard~\cite{IEEE8023}. 
A dedicated firmware project was developed to evaluate the performance of both the state machine and the UDP/IP stack.

The test design consists of an instance of the \texttt{mgt\_to\_axi} module connected to the 10G UDP stack.
A test harness injects data into the \texttt{mgt\_to\_axi} input FIFO.
Two 32-bit counters are implemented: a \emph{cycle counter}, incremented every clock cycle to measure elapsed time, and a \emph{word counter}, incremented only when the input FIFO is not full.
The word counter output is streamed into the data path and received on a remote server.

The measurement starts upon the first assertion of the UDP stack \texttt{ready} signal and stops when the cycle counter reaches \texttt{0x7fffffff} (\texttt{positive'high}), corresponding to approximately 13~s of acquisition.
This duration exceeds the typical \TimepixFour\ test-beam spill length (5--6~s), ensuring steady-state conditions.

The maximum sustainable zero-back-pressure data rate is estimated as:

$$ rate \mathrm{\left[ \frac{bits}{s} \right]} = \frac{(word~counter~value) \times 64~\mathrm{[bits/word]}}{6.4\cdot 10^{-9}~\mathrm{[s/cycle]} \times (cycle~counter~value)}$$

Here the $64~\mathrm{[bits/word]}$ factor represents the ethernet clock period and $64~\mathrm{[bits/word]}$ is the numer of bits per transmitted word.
A sustained data rate of $8.85\pm 0.05~\mathrm{Gb/s}$ was measured, consistent with the theoretical peak UDP payload limit of approximately $\approx 9.5~\mathrm{Gb/s}$ for standard MTU frames.
The difference is attributed to packetization overhead and implementation inefficiencies.
This performance satisfies the requirements of the present project.

\textbf{Transmission reliability} - To test data transmission reliability a PRBS checker is included in the FPGA transceiver instantiation. This PRBS checker can be enabled and read back using a set of \ipbus registers. A PRBS test is performed by setting the \TimepixFour output links to transmit a PRBS-31 sequence. The checker on the FPGA receiver is enabled. A simple counter is used to keep track of the errors occourred over a period of 10s. This counter is then read-back via the control path. This test performed simultaneously on all the \TimepixFour links, and it is used to select a link speed and MGT redriver configuration for which no data corruption is registered for all links. This test shows that reliable data tacking is achievable using line rates of up to $2.56\ \mathrm{Gb/s}$ using the \FourDPhoton Tube card (measured Bit Error Rate $< 3.9\cdot 10^{-11}$). 

\textbf{Slow control interface transaction rate} - Although less important, performances of the slow control have been evaluated.
The performance of this protocol can greatly affect the configuration time of the \gls{asic}.
Furthermore if communication on the multi gigabit lines can not be guaranteed this interface can be used to readout the ASIC thus good performance is required.
The performance where evaluated in a real live scenario by subsequently polling a single 256-bit long \TimepixFour register.
This register is chosen to test the most critical operation, that is data readout.
The performed tests indicate a transaction rate of $13\pm0.5~\mathrm{KOps/s}$ is achievable.

\subsection{System synchronization performance}

The synchronization performance of the system was evaluated in experimental conditions using two independent \TimepixFour\ ASICs connected to two separate copies of the \gls{idaq} system. 
Both systems received the same reference clock and synchronization signals (\texttt{T0\_SYNC} and \texttt{SHUTTER}) generated by the \gls{tlu}. 
A common test pulse was simultaneously injected into the two ASICs to evaluate the timing alignment between the acquisition chains.

The test pulse was generated using an Active Technologies AWG4018 arbitrary waveform generator triggered by a copy of the \texttt{SHUTTER} signal distributed by the \gls{tlu} (see Section~\ref{sec:tlu}). 
According to the manufacturer specifications, the channel skew of the generator is below $20~\mathrm{ps}$.

\begin{figure}[htb]
\centering
\includegraphics[width=0.75\textwidth]{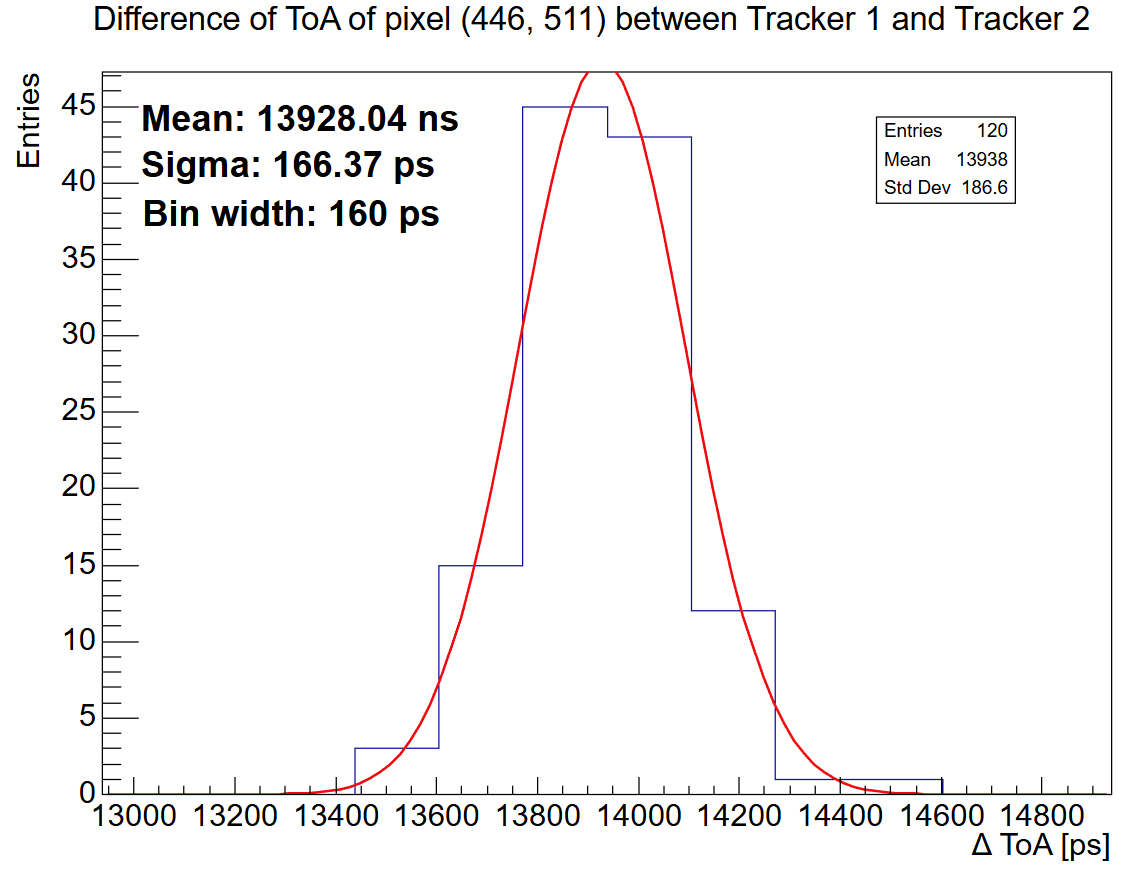}
\caption{Timing resolution of the dual-\gls{idaq} system.}
\label{fig:timing_resolution}
\end{figure}

Synchronization performance was evaluated by comparing the timestamps recorded by two ASICs for coincident events as shown in Figure~\ref{fig:timing_resolution}.
The third \TimepixFour ASIC represented the device under test for the experiment and is not considered in this analysis.
The timestamp distribution was fitted with a Gaussian function and the standard deviation $\sigma$ of the distribution was extracted. 
Since the measured spread includes contributions from the waveform generator skew and from the intrinsic timestamp quantization of the two ASICs, the synchronization resolution was estimated as:

\[
\mathrm{res} =
\sqrt{
\frac{\sigma^2
-
(20~\mathrm{ps})^2
-
(72~\mathrm{ps})^2
-
(72~\mathrm{ps})^2
}{2}} = 92~\mathrm{ps}
\]

where the $20~\mathrm{ps}$ term accounts for the channel to channel skew of the waveform generator for the active channels, the two $72~\mathrm{ps}$ terms account for the timing resolution contribution of the two \TimepixFour\ ASICs~\cite{Bolzonella2024} and the $\sqrt{2}$ factor derives from the use of two \gls{idaq} systems in parallel.

The measured synchronization resolution of $92~\mathrm{ps}$ demonstrates deterministic timing alignment between multiple \gls{idaq} systems operating from a common clock and trigger distribution network. 
This result was obtained under experimental conditions during a beam-test campaign and therefore represents an upper bound on the intrinsic synchronization precision of the system.

\section{Conclusions}\label{sec:conclusion}

A flexible and modular \gls{daq} system for the \TimepixFour\ \gls{asic} has been developed within the \FourDPhoton project. 
The system comprises the complete acquisition chain, a multi-board synchronization mechanism, and a dedicated trigger logic unit. 
It has been designed to support full characterization of the \FourDPhoton detector and dedicated beam-test campaigns.

The achieved performance enables complete detector testing and has been validated through device acceptance tests and beam-test measurements. 
System throughput and timing performance are primarily determined by the hardware architecture; further improvements would require custom hardware or application-specific optimizations.

The \gls{idaq} system supports scalable multi-board operation for beam-test applications. 
Synchronization across up to three boards has been demonstrated and the same mechanism can be applied for up to 6 boards with existing hardware.

\section{Outlook and Future Developments}

Further developments will focus on improving portability and long-term maintainability of the firmware. 
In particular, vendor-specific IP cores will progressively be replaced by open-source logic libraries. 
Memory controllers, clock-domain crossing (CDC) modules, and related infrastructure blocks will migrate toward reusable open-source components such as Open Logic~\cite{OpenLogic}. 
Similarly, the current vendor-provided 1~GbE MAC will be replaced with an implementation from the \textit{verilog-ethernet} library~\cite{VerilogEthernet}. 
These changes will ease migration across FPGA families and reduce toolchain dependencies.

An extension of the described system to support the \MedipixFour  \gls{asic} is under development.
This extension will exploit the same the control path as the described system while implementing a dedicated data path. 

A port of the system to the \textit{Alinx AXKU040} development board is also foreseen. 
This board integrates the same Kintex UltraScale XCKU040 FPGA and provides an identical number of FMC transceivers, ensuring firmware compatibility, while offering four 10~GbE interfaces and therefore increased maximum output bandwidth. 
Peripheral adaptations, such as changes in the 1~GbE PHY interface, are required but do not impact the core firmware architecture.

\acknowledgments

This work was carried out in the context of the Medipix4 Collaboration based at CERN, and in the framework of the MEDIPIX4 project funded by INFN CSN5. This project has received funding from the European Research Council (ERC) under the European Union’s Horizon 2020 research and innovation programme (Grant agreement No. 819627, 4DPHOTON project).

\bibliographystyle{iopart-num}

\end{document}